\documentclass[11pt,preprintnumbers,aps,amssymb,nofootinbib,amsmath,superscriptaddress,notitlepage]
{revtex4-1}
\usepackage{epsfig,epsf}
\usepackage{bm} % puts greek and math symbols in boldface using \bm
\usepackage{color} % {\color{red} ... }
\usepackage{slashed}
\usepackage{relsize}	%rescalable math 
\usepackage{soul} %spac­ing out (let­terspac­ing), un­der­lin­ing, strik­ing out, etc.
\usepackage{hyperref}
\usepackage{tensor} %\indices{^a_b} produces the properly spaced the tensor indices 
\newcommand{\beq}{\begin{equation}}
\newcommand{\beql}[1]{\begin{equation}\label{#1}}
\newcommand{\eeq}{\end{equation}}
\def\bal#1\gal{\begin{align}#1\end{align}}
\newcommand{\ball}[1]{\bal\label{#1}}
\newcommand{\eq}[1]{(\ref{#1})}
\newcommand{\fig}[1]{Fig.~\ref{#1}}
\renewcommand{\sec}[1]{Sec.~\ref{#1}}
\renewcommand{\b}[1]{{\bm #1}} 
\newcommand{\unit}[1]{\hat {{\bm #1}}} % unit vector

\newcommand{\jpsi}{J\mskip -2mu/\mskip -0.5mu\psi} % J/Psi with proper spacing

\begin{document}

%\preprint{RBRC-814}

\title{Bound states and electromagnetic radiation of relativistically rotating cylindrical wells }

\author{Matteo Buzzegoli}

\author{Kirill Tuchin}

\affiliation{
Department of Physics and Astronomy, Iowa State University, Ames, Iowa, 50011, USA}

\date{\today}

\begin{abstract}

We compute the effect of rigid rotation on the non-relativistic bound states. The energy levels of the bound states increase with the angular velocity of rotation until at certain value of the angular velocity they are completely pushed out into the continuum which corresponds to dissociation of the bound states. When the angular velocity exceeds the critical value at which the ground state disappears into the continuum, no bound state is possible. This effect should have important consequences for the phenomenology of the quark-gluon plasma. One of the ways to study it experimentally is to observe the electromagnetic radiation emitted by a rotating bound state. We compute the corresponding  intensity of electromagnetic radiation and show that it strongly depends on the angular velocity of rotation.    

\end{abstract}

\maketitle

%%%%%%%%%%%%%%%%%%%%%%%%%%%%%%%%%%%%%%%%
\section{Introduction}\label{sec:a}

The study of the properties of rotating systems has a long history. Relatively recent studies focused on thermodynamics and kinetic of rotating quantum systems \cite{Vilenkin:1980zv,Becattini:2007nd,Ambrus:2014uqa,McInnes:2014haa,McInnes:2015kec,Ambrus:2015lfr,Chen:2015hfc,McInnes:2016dwk,Jiang:2016wvv,Chernodub:2016kxh,Ebihara:2016fwa,Chernodub:2017ref,Liu:2017spl,Buzzegoli:2017cqy,Zhang:2018ome,Buzzegoli:2020ycf,Chen:2019tcp,Wang:2018zrn,Wang:2019nhd,Palermo:2021hlf,Sadooghi:2021upd}. A great interest in relativistic rotating systems was recently spurred by the phenomenology of the relativistic heavy-ion collisions that indicates that the quark-gluon plasma has large local vorticity \cite{STAR:2017ckg,Csernai:2013bqa,Deng:2016gyh,Jiang:2016woz,Xia:2018tes}. In fact, the vorticity $\Omega$ found in numerical simulations  approaches the relativistic limit $\Omega r=1$, where $r$ is the radial distance from the rotation axis. The importance of the causal boundary was emphasized by a number of authors \cite{Duffy:2002ss,Ebihara:2016fwa,Chernodub:2017ref,Chernodub:2017mvp}.

One of the plasma signatures is dissociation of the bound states immersed into it through the mechanism of the Debye screening. The ability of plasma to break up the bound states depends on the relationship between the Debye radius which is a decreasing function of temperature and the linear size of the bound state. On the other hand, the rotating plasma drags the bound state along. As a result, the rotating bound state possesses extra centrifugal energy that makes it more fragile. Thus, both the Debye screening and rotation induce the break up of bound states. The main goal of this paper is to investigate the effect of rotation on the bound state spectrum. 

To this end we consider a model consisting of a cylindrical well of radius $R$ and constant depth $U_0$, rotating with constant angular velocity $\Omega$ about an axis passing through its center-of-mass. We neglect all plasma effects apart from being the source of rotation. Furthermore, we treat the bound state non-relativistically which allows us to retain essential qualitative features while keeping algebra simple. Our results thus apply literally to a heavy quarkonium immersed into and coaxial with a vortex of the rotating plasma.  The critical ingredient of our calculation is the causal boundary condition at $r=1/\Omega$. The results strongly depend on whether $\Omega R$ is larger or smaller than unity. We refer to this two cases as rapid and slow rotation. Our main observation is that the energy levels of the rotating quarkonium increase with $\Omega$ and get pushed out of well at a certain $\Omega$ that is larger for deeper wells. We study these matters in Sec.~\ref{sec:rot-bound}.
   
The dependence of the spectrum on $\Omega$ affects the electromagnetic radiation emitted by the quarkonium. In particular, we found that when $\Omega>1/R$ the intensity increases by $(\Omega R)^6$ in comparison with the non-rotating well. This is studied it Sec~\ref{sec:radiation}. This means that the life-time of the excited quarkonium states are sharply decrease with $\Omega$. Similar conclusion can be made with regard to the gluon strong and electroweak interactions as well. This observation may be instrumental for the experimental investigation of the rotation effects on the bound states.

%%%%%%%%%%%%%%%%%%%%%%%%%%%%%%%%%%%%%%%%
\section{Bound states of stationary cylindrical well}\label{sec:b}

\begin{figure}[t]
\begin{tabular}{cc}
      \includegraphics[height=5cm]{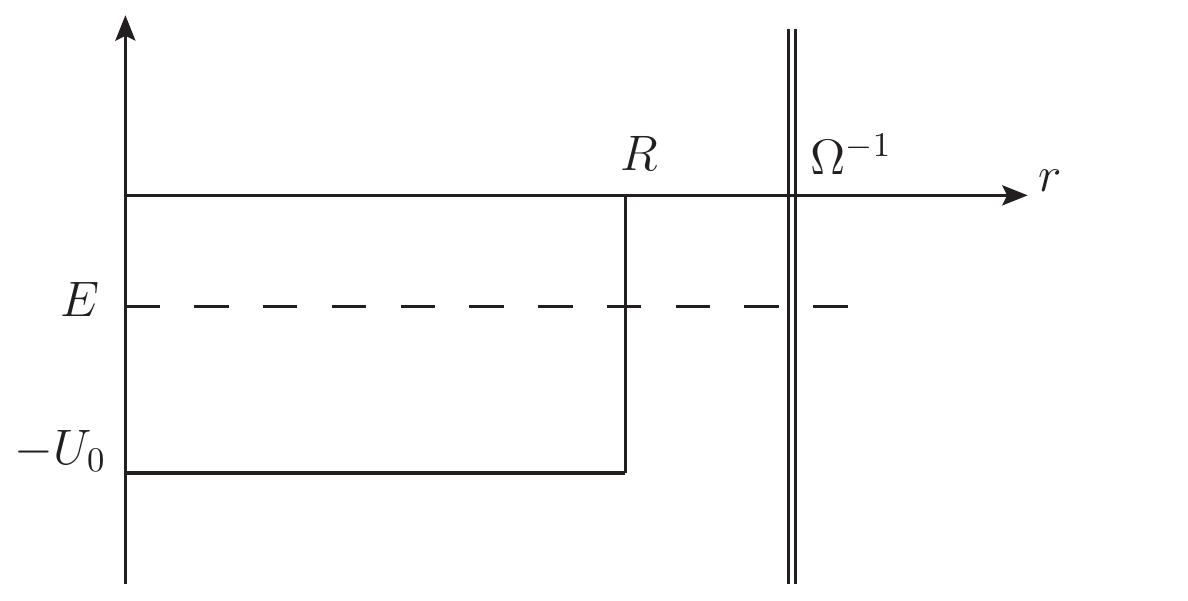} &
      \includegraphics[height=5cm]{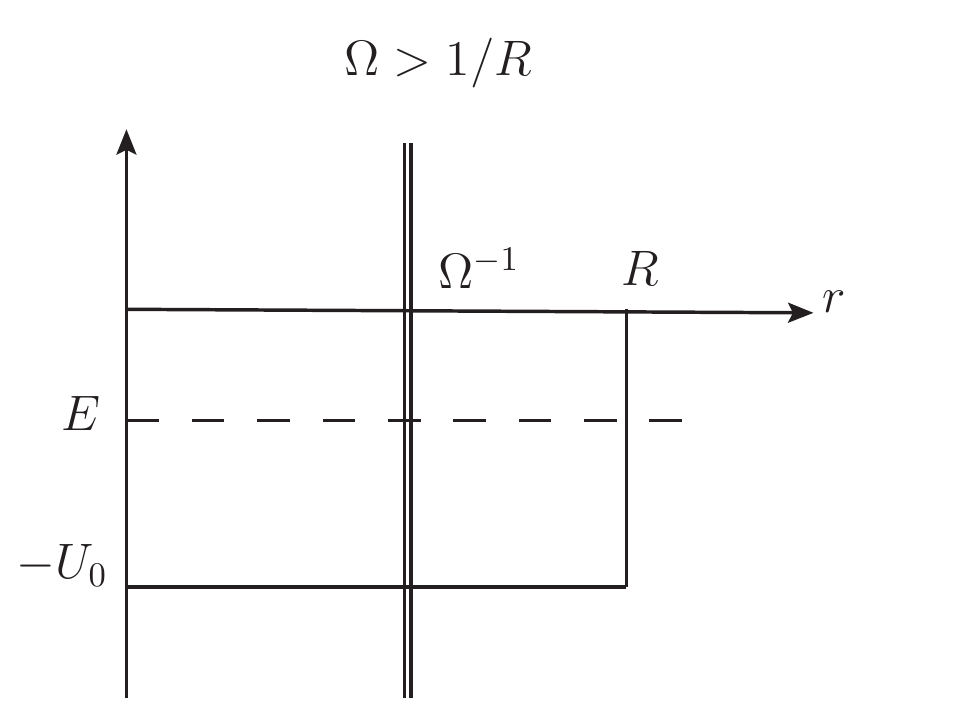}
      \end{tabular}
  \caption{ Potential cylindrical well rotating with angular velocity $\Omega$. Because of the causality, the wave function must vanish at $r>1/\Omega$.  Left panel: slow rotation, right panel: fast rotation. In the stationary case $\Omega^{-1}\to \infty$. }
\label{fig:potential}
\end{figure}

First, for the future reference, consider a non-rotating cylindrical well as shown in \fig{fig:potential}. We are interested in the solutions with $E<0$.  The Schr\"odonger equation inside the well $r< R$, reads 
\ball{b3}
H_0\psi_0= -\frac{1}{2M}\nabla^2\psi_0 - U_0\psi_0=E_0\psi_0\,.
\gal
The corresponding set of eigenfunctions, regular at the origin and  bound in the radial direction, is 
\ball{b5}
\psi_0 (\b x)=AJ_m(kr) e^{im\phi}e^{ik_zz}\,, 
\gal
where $J_m(z)$ is the Bessel function of the first kind, $A$ is the normalization constant and 
\ball{b7}
k=\sqrt{2M(U_0-k_z^2/2M-|E_0|)}\,.
\gal
Clearly the bound states exist only if $|E_0|\le U_0-k_z^2/2M$.

Outside the well the functions that vanish at $r\to \infty$ are 
\ball{b11}
\psi_0 (\b x)=BK_m(\kappa r) e^{im\phi}e^{ik_zz}\,, 
\gal
where $K_m(z)$ is the modified Bessel function of the second kind and
\ball{b13}
\kappa= \sqrt{2M(k_z^2/2M+|E_0|)}\,.
\gal
Eq.~\eq{b11} satisfy the boundary condition $\psi\to 0$ as $r\to \infty$. 

The boundary conditions at $r=R$ are the requirements of continuity of the wave function and its first derivative:
\bal
A J_m(kR)&= BK_m(\kappa R)\,,\label{b16}\\
A J_m'(kR)k&= BK_m'(\kappa R)\kappa\,.\label{b17}
\gal
The solution exists only if the determinant vanishes:
\ball{b19}
J_m(kR)K_m'(\kappa R)\kappa=K_m(\kappa R)J_m'(kR)k \, .
\gal
This equation along with 
\ball{b21}
k=\sqrt{2MU_0-\kappa^2}
\gal
 gives the spectrum.

%%%%%%%%%%%%%%%%%%%%%%%%%%%%%%%%%%%%%%%%
\section{Bound states of rotating cylindrical well}\label{sec:rot-bound} 

The Hamiltonian of the rotating system  is $H=H_0-\Omega J_z$. Thus, in cylindrical coordinates, $H\psi= E\psi$ is equivalent to $H_0\psi= (E+m\Omega)\psi$, where $m$ is an integer eigenvalue of the operator $J_z=-i\partial_\phi$.  The general solution of the latter equation was found in the previous section, we just need to substitute $E_0= E+m\Omega$ and apply the new boundary conditions. We proceed by considering separately the slowly and rapidly rotating well.

%%%%
%%%%%

%%
\subsection{Slow rotation $R<1/\Omega$}\label{sec:c}

The wave functions inside the well have the same functional form as \eq{b5}. However, $k$ is now given by 
\ball{c7}
k=\sqrt{2M(U_0-k_z^2/2M-|E+m\Omega|)}\,, \qquad E+m\Omega<0\,.
\gal

The solution outside the well reads
\ball{c11}
\psi (\b x)=\left[ BK_m(\kappa r)+CI_m(\kappa r)\right] e^{im\phi}e^{ik_zz}\,, 
\gal
where $I_m(z)$ is the modified Bessel function of the first kind and
\ball{c13}
\kappa= \sqrt{2M(k_z^2/2M+|E+m\Omega|)}\,.
\gal

Unlike $\psi_0$ in \eq{b11}, eigenfucntions \eq{c11} cannot extend to infinity. Instead, causality requires that $\psi$ vanishes at $r=1/\Omega$:
\ball{c15}
 BK_m(\kappa /\Omega)+CI_m(\kappa /\Omega)=0\,.
 \gal
Additionally, the boundary conditions at $r=R$ yield
\bal
A J_m(kR)&= BK_m(\kappa R)+C I_m(\kappa R)\,,\label{c16}\\
A J_m'(kR)k&= BK_m'(\kappa R)+C I_m'(\kappa R)\kappa\,.\label{c17}
\gal
The determinant of the set of linear equations \eq{c15}, \eq{c16}, \eq{c17} must vanish:
\ball{c19}
I_m&(\kappa/\Omega) \left[-J_m(kR)K_m'(\kappa R)\kappa+K_m(\kappa R)J_m'(kR)k\right]\nonumber\\
&-K_m(\kappa/\Omega)\left[ -J_m(kR)I_m'(\kappa R)\kappa+I_m(\kappa R)J_m'(kR)k\right]=0 .
\gal
As $\Omega\to 0$ this equation reduces to \eq{b19}.

Using the convenient  notation: $x=kR$, $y=\kappa R$, $\lambda =R\Omega$, $\beta=\sqrt{2MU_0R^2}$,
we can write the boundary condition \eq{c19} as: 
\bal
I_m&(y/\lambda) \left[-J_m(x)K_m'(y)y+K_m(y)J_m'(x)x\right]\nonumber\\
&-K_m(y/\lambda)\left[ -J_m(x)I_m'(y)y+I_m(y)J_m'(x)x\right]=0\,, \label{c21}
\gal 
where $x$ and $y$ are related to each other as 
\bal
&x=\sqrt{\beta^2-y^2}\,.\label{c22}
\gal
Denoting the roots of \eq{c21} as  $y_{mn}$, $n=1,2,\ldots$ we obtain the energy spectrum 
\ball{c23}
E_{mnk_z}= -\frac{1}{2MR^2}y_{mn}^2-\frac{m\lambda}{R}+\frac{k_z^2}{2M}\,.
\gal

%%%%
\begin{figure}[t]
      \includegraphics[height=5cm]{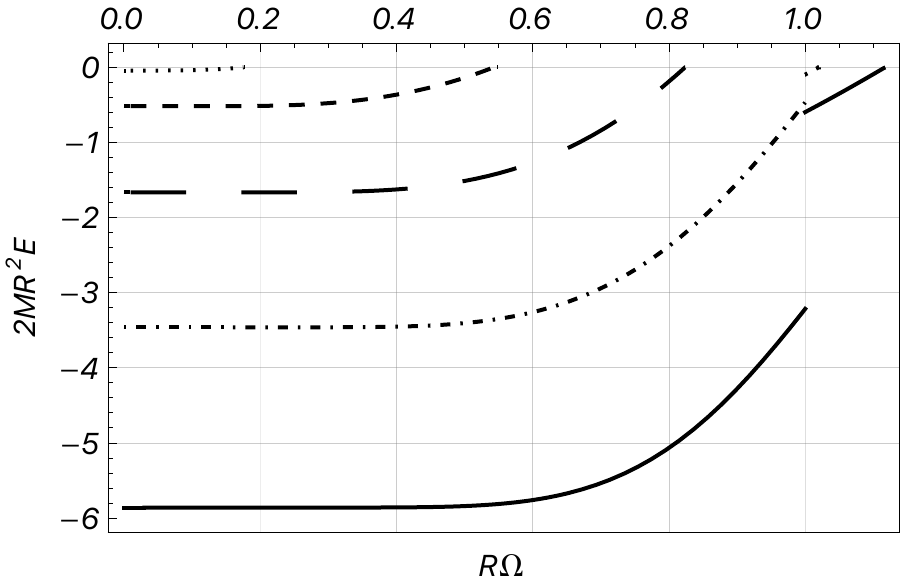} 
  \caption{Energy levels \eq{c23} and \eq{d6} with $m=0$, $k_z=0$ in wells of different depth: $\beta=1$ (dotted line), $\beta=1.5$ (short-dashed line), $\beta=2$ (long-dashed line), $\beta= 2.5$ (dashed-dotted line) and $\beta=3$ (solid line) as a function of  $\lambda=R\Omega$.  Notice the discontinuity at $\lambda=1$.  }
\label{fig:slow}
\end{figure}
%%%%%

%%%%
%\begin{figure}[t]
%\begin{tabular}{cc}
 %     \includegraphics[height=4.5cm]{slow.pdf} &
 %     \includegraphics[height=4.5cm]{slow2.pdf}
 %     \end{tabular}
%  \caption{ Energy level $m=0$, $k_z=0$ in the well with $\beta=1$ (blue), $\beta=1.2$ (red), $\beta=2.4$ (brown), $\beta= 2.5$ (black) and $\beta=3$ (purple) as a function of angular velocity $\Omega$.  Notice the discontinuity at $\lambda=1$.  }
%\label{fig:slow}
%\end{figure}
%%%%%

%%%%%%%%%%%%%%%%%%%%%%%%%%%%%%%%%%%%%%%%
\subsection{Rapid rotation}\label{sec:d}

If the rotation is fast $R>1/\Omega$, all the region $r>1/\Omega$ is outside of the rotating spacetime. Thus there is only solution inside the well; its functional form is \eq{b5} and the boundary condition is
\ball{d3}
J_m(k/\Omega)=0\,.
\gal
Denoting  zeros of the $m$'th Bessel function as $x_{ma}$, $a=1,2,\ldots$  we obtain $k=\Omega x_{ma}$. The spectrum of a particle in the rapidly rotating cylindrical well is
\ball{d5}
E_{mak_z}= \frac{1}{2M}(\Omega^2 x^2_{ma}+k_z^2)-U_0-m\Omega\,,
\gal
or, equivalently, 
\ball{d6}
E_{mak_z}= \frac{1}{2MR^2}\left( \lambda^2 x^2_{ma}-\beta^2\right) +\frac{k_z^2}{2M}-\frac{m\lambda}{R}\, .
\gal

\fig{fig:slow} displays the energy level \eq{c23}, for $\lambda<1$, and \eq{d6}, for $\lambda>1$, with $m=0$, $k_z=0$ in the potential wells of different depths. 
As the angular velocity of rotation increases, the level increases (i.e.\ its absolute value decreases) until it reaches $E=0$.  The discontinuity at $R\Omega=1$ is an artifact of our model and disappears when the boundary is treated more carefully, see \sec{sec:coulomb}. For the range of $\beta$ shown in the figure, there is only one level, but in a deeper well there are more energy levels. All of them exhibit similar dependence on $\Omega$.

%%%
\subsection{Shallow and slowly rotating cylindrical well}\label{sec:e}

Consider states with $m=0$, $k_z=0$ in the shallow well limit $\beta^2\ll 1$,  we can expand the Bessel functions in \eq{c21} at  $x, y\ll 1$ which yields
\ball{e3}
1-\frac{1}{2}(\beta^2-y^2)\ln\frac{2}{\tilde\gamma y}= -\frac{\pi\beta^2 }{2}\exp\left(-\frac{2y}{\lambda}\right)\,,
\gal
where $\tilde \gamma=\exp \gamma \approx 1.78$, with $\gamma$ being Euler-Mascheroni constant.
Considering the right-hand-side of \eq{e3} a perturbation we can write at the leading ``non-rotating'' approximation
 \ball{e5}
1-\frac{1}{2}(\beta^2-y_0^2)\ln\frac{2}{\tilde \gamma y_0}=0\,.
\gal
The only solution is obtained by neglecting $y_0$ compared to $\beta$:
\ball{e7}
y_0=\frac{2}{\tilde\gamma}\exp\left(-\frac{2}{\beta^2}\right)\,.
\gal
Clearly, $y_0$ is much smaller than any positive power of (small) $\beta$. In usual units
\ball{e8}
E_0=-\frac{2}{\tilde \gamma^2 MR^2}\exp\left(-\frac{2}{MU_0R^2}\right)\,.
\gal

The second iteration of \eq{e3} is obtained by expanding $y=y_0+y_1$ at small $y_1\ll y_0$:
\ball{e9}
y_1= -\pi y_0\exp\left(-\frac{4}{\lambda\tilde \gamma}e^{-\frac{2}{\beta^2}}\right)\,.
\gal
Converting this back to the dimensional quantities gives the correction due to  rotation
\ball{e10}
\Delta E=-2\pi E_0 \exp\left(-\frac{4}{R\Omega\tilde \gamma}e^{-\frac{2}{MU_0R^2}}\right)\,.
\gal
As $\Omega$ increases, $\Delta E$ also increases with the result that the energy level moves up.  

The level crosses into the continuum when $E_0+\Delta E=0$. This happens at the value $\Omega_c$ given by 
\ball{e12}
\Omega_c=\frac{2\sqrt{2M|E_0|}}{\ln(2\pi)}\,.
\gal

%%%%%%%%%%%%%%%%%%%%%%%%%%%%%
\section{Bound states of rotating 2D $1/r$ potential}\label{sec:coulomb}

Another instructive model of a rotating well with cylindrical symmetry is an attractive $U=-\alpha/r$ potential in two dimensions. Unlike the previous example, this potential does not have a sharp boundary and hence no discontinuity of the energy levels observed in \fig{fig:slow}.

As in the previous section we first consider the stationary problem
\ball{g3}
\left( \b\nabla^2 + \frac{2M\alpha}{r}-2M|E_0|\right)\psi_0=0\,.
\gal
Its solution, finite at the origin, is
\ball{g15} 
\psi_0= A\, _1F_1(-n+|m|+1/2,2|m|+1,\rho)\rho^{|m|}e^{-\rho/2}e^{im\phi}\,,
\gal
where $\,_1F_1$ is the confluent hypergeometric function of the first kind and we defined
\ball{g17}
\rho =\frac{2\alpha M}{n}r\,,\qquad n= \frac{1}{\sqrt{2|E|}}\,.
\gal
Requiring that at $r\to \infty $ the wave function \eq{g15} is finite gives the quantization condition $n=n'-1/2$ with $n'=1,2,\ldots$ and $|m|\le n'-1$.

The boundary condition of the  rotating potential requires vanishing of the wave function at $r=1/\Omega$, i.e.\,
\ball{g21}
\psi\big|_{\rho = 2\ell/n}=0\,,\qquad \ell = \frac{\alpha M}{\Omega}\,.
\gal 
As a result $n'$ depends on $\Omega$ and is not integer anymore. Solving \eq{g21} for the given values of $m$ and $\ell$  we  find the spectrum of $n'$. The energy levels are then given by
\ball{g23}
E_{n',m}= -\frac{\alpha^2M}{2(n'(\Omega)-1/2)^2}-m\Omega\,.
\gal

\begin{figure}[t]
\begin{tabular}{lr}
      \includegraphics[height=6cm]{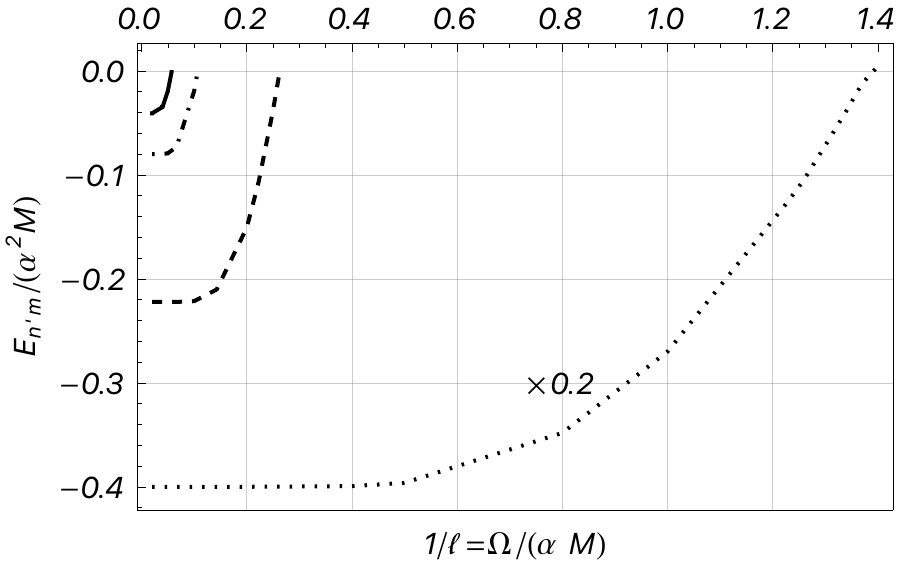} &
      \includegraphics[height=6cm]{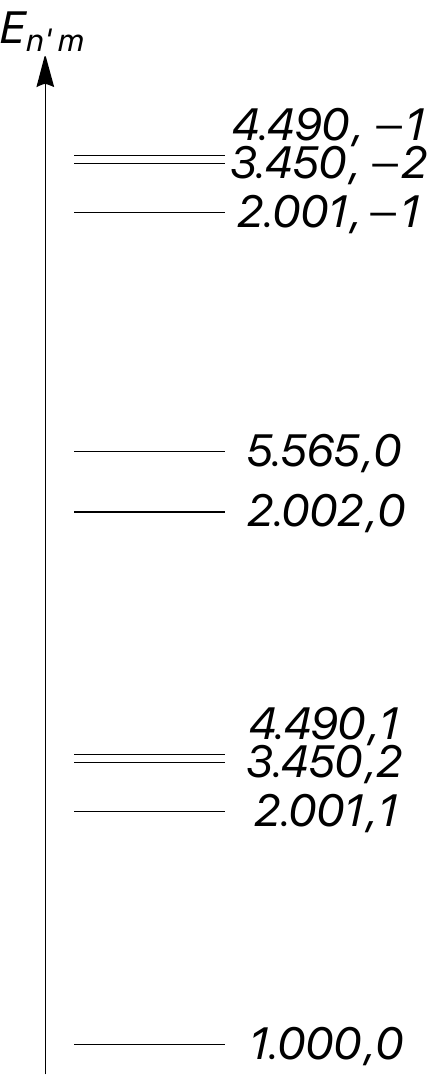}
      \end{tabular}
  \caption{  Left panel: Energy levels $m=0$, $n'=1,2,3,4$ (dotted, dashed, dashed-dotted and solid lines respectively) in the two-dimensional potential $U=-\alpha/r$ at different values of $1/\ell$.   Right panel: energy levels with given $n'$ (first number) and $m$ (second number) at $\ell=10$ and $\alpha \ell =1$.}
\label{fig:2dcoulomb}
\end{figure}

\fig{fig:2dcoulomb} (left panel) shows $E_{n',0}$ as a function of $\ell$. The qualitative behavior is similar to \fig{fig:slow}. As the angular velocity of rotation $\Omega$  increases the energy levels get pushed out of the well until at $\ell=\ell_{1,0}=0.72$ the last azimuthally symmetric level ($m=0, n'=1$)\footnote{More precisely, the level that has $m=0$, $n'=1$ at $\Omega=0$.} disappears indicating absence of such bound states. The next azimuthally symmetric level ($m=0, n'=2$) disappears at $\ell=\ell_{2,0}=3.84$. 

%%%%
%%%%%

%%%%%%%%%%%%%%%%%%%%%%%%%%%%%
\section{Electromagnetic radiation by rotating cylindrical well}\label{sec:radiation}

A rotating well binding oppositely charged particles emits electromagnetic radiation. The intensity of radiation can be used as an effective diagnostic tool. 

The $S$-matrix element describing photon radiation when the system transitions from the initial state $i$ to the final state $f$ is
\ball{f1}
S_{fi}=-ie\int dt\int d^3x\, \b j_{fi}(x)\cdot \b A^*(x)\,,
\gal
where the transition current is
\ball{f2}
\b j_{fi}= \frac{i}{2M}\left( \psi_i\b\nabla \psi^*_f-\psi^*_f\b\nabla \psi_i\right)\,.
\gal
We will distinguish the quantum numbers of the final state with the prime, i.e.\ $i=\{m,k_z,\ldots\}$, $f=\{m',k_z',\ldots\}$. 
The well wave functions, including their time dependence, read
\ball{f5}
\psi (x)=\frac{1}{2\pi}X(r) e^{im\phi}e^{ik_zz}e^{-iEt}\,, 
\gal
where $X(r)$ is an appropriate radial function derived in the previous section. The wave functions are normalized by $\int \psi^*\psi d^3 x=1$ which implies
\ball{f6}
\int_0^{1/\Omega} dr r X^2=1\,.
\gal

 The photon wave function is
\ball{f7}
\b A^*(x)= \frac{1}{\sqrt{2\omega V}}\b\epsilon^*_{\b p,\lambda} e^{-i\b p\cdot \b x+i\omega t}\,,
\gal
where $\b\epsilon^*_{\b p,\lambda}$ is the polarization vector, $\lambda=\pm 1$. 

Since we treat the well in non-relativistic approximation, the leading contribution to the radiation intensity is the electric dipole radiation corresponding to wave lengths much larger than the system size, i.e. $pR\ll 1$. Thus neglecting the phase $\b p\cdot\b r$ in the exponent in \eq{f7} and substituting it along with \eq{f5} into \eq{f2} and \eq{f1}  produces 
\ball{f9}
S_{fi}=\frac{e(2\pi)^2}{2M\sqrt{2\omega V}}\b\epsilon^*_{\b p,\lambda}\cdot  \b I_{fi}
\delta(E-E' -\omega)\delta_{mm'}\delta_{k_zk_z'}\,,
\gal
where we defined 
\ball{f10}
\b I_{fi}&= I_{r,fi}\unit r+ I_{\phi,fi}\unit \phi
+I_{z,fi}\unit z\,\nonumber\\
&
= \unit r\int_0^{1/\Omega} (X_f'X_i-X_fX_i')rdr
+2m\unit \phi \int_0^{1/\Omega} X_f X_i dr
+ 2k_z\unit z\int_0^{1/\Omega}X_f X_i r dr \,.
\gal
The total radiation intensity is given by dividing the emission probability $|S|^2$ by the observation time $\Delta t$, multiplying by the photon energy $\omega$ and summing and integrating  over the photon phase space:
\ball{f12} 
W_{fi}= \sum_\lambda \int \frac{|S|^2}{\Delta t}\omega\frac{do_{\unit p}\omega^2d\omega V}{(2\pi)^3}\,,
\gal
where $do_{\unit p}$ is the solid angle element in the direction of the photon momentum. Substituting \eq{f12} into \eq{f9} and using 
\ball{f14}
\sum_\lambda |\b\epsilon^*_{\b p,\lambda}\cdot \b I_{fi} |^2= \frac{|\b I_{fi} \times \b p|^2}{\omega^2}\,,
\gal
we obtain
\ball{f17}
W_{fi}= \frac{1}{24\pi}\frac{e^2\omega^2}{M^2}|\b I_{fi}|^2\delta_{mm'}\delta_{k_zk_z'}\,.
\gal

Eqs.~\eq{f17} and \eq{f10} give the intensity of the electromagnetic radiation in the dipole approximation. The photon energy $\omega$ is fixed by the energy and momentum conservation and depends on the angular velocity $\Omega$.

%%%%%%%%%%%
\subsection{Radiation by rapidly rotating cylindrical well}

Using the model we developed in the previous section we can compute the intensity of electromagnetic radiation of the rapidly rotating cylindrical well $R>1/\Omega$. For simplicity we consider a reference frame where $k_z=0$. The radial wave functions normalized by \eq{f6} and satisfying the boundary condition \eq{d3} are
\ball{f20}
X(r)=\frac{\sqrt{2}\Omega}{J_{m+1}(x_{ma})}J_m(x_{ma}\Omega r)\,.
\gal
Using these in \eq{f10} and plugging into \eq{f17} furnishes 
\ball{f22}
W_{fi}= \frac{1}{96\pi}\frac{e^2\Omega^6}{M^4}(x_{ma}^2-x_{ma'}^2)^2
\Omega^2(\tilde I_{r,fi}^2+\tilde I_{\phi,fi}^2)
\delta_{mm'}\,,
\gal
where the photon energy is, using \eq{d6}, 
\ball{f24}
\omega = E_{ma0}-E_{ma'0}= \frac{\Omega^2}{2M}(x_{ma}^2-x_{ma'}^2)\,,
\gal
and the functions 
\bal
\tilde I_{r,fi}&= \int_0^1 \frac{2\rho d\rho }{J_{m+1}(x_{ma})J_{m+1}(x_{ma'})}\left[
J_m'(x_{ma'}\rho)J_m(x_{ma}\rho)x_{ma'}-J_m(x_{ma'}\rho)J_m'(x_{ma}\rho)x_{ma}\right],\\
\tilde I_{\phi, fi}&=\int_0^1 \frac{4m d\rho }{J_{m+1}(x_{ma})J_{m+1}(x_{ma'})}
J_m(x_{ma'}\rho)J_m(x_{ma}\rho)\,
\gal
are independent of $\Omega$. They can be expressed in terms of the hypergeometric function.  

%%%%%%%%%%%
\subsection{Radiation by rotating infinite cylindrical well}

The calculation of the radiation intensity for slowly rotating cylindrical well $R<1/\Omega$ can also be done analytically. However it leads to bulky expressions involving Bessel functions. A more instructive model is an \emph{infinite} cylindrical well, i.e.\ the potential $U=0$ if $r\le R$ and $U=\infty$ otherwise. The corresponding wave functions are similar to \eq{f20}
\ball{f30}
X(r)=\frac{\sqrt{2}}{\ell J_{m+1}(x_{ma})}J_m(x_{ma} r/\ell)\,, 
\gal
where $\ell = R$ if $R<1/\Omega$ (slow rotation) and $\ell =1/\Omega$ otherwise. The intensity is now given by (see \eq{f22}):
\ball{f32}
W_{fi}= \frac{1}{96\pi}\frac{e^2}{M^4\ell^6}(x_{ma}^2-x_{ma'}^2)^2(\tilde I_{r,fi}^2+\tilde I_{\phi,fi}^2)\delta_{mm'}\,.
\gal
Thus, in intensity is $\Omega$ independent for slow rotation. However, once $\Omega>1/R$ the intensity increases rapidly in proportion to $\Omega^6$:
\ball{f34}
W_{fi}= (R\Omega)^6W_{fi}^{\Omega=0}\,, \qquad \text{if}\quad \Omega R>1\,.
\gal
While the particular functional form of $W_{fi}(\Omega)$ depends on the form of the potential $U$, the observation of the steep growth of the radiation intensity with $\Omega$ appears to be quite general.

%%%%%%%%%%%%%%%%%%%%%%%%%%%%
\section{Discussion}\label{sec:sum}

We computed the energy spectrum of a particle in the non-relativistic cylindrical potential well rotating with constant angular velocity $\Omega$. The energy levels increase with the angular velocity $\Omega$ until they are pushed out of the well at some critical value $\Omega_c$ as can be seen in \fig{fig:slow}. The particular form of the functional dependence $E(\Omega)$ is model-dependent. The discontinuity at $R\Omega=1$ is an artifact of the cylindrical well approximation; it disappears for a smoother potential such as the attractive $1/r$ potential discussed in \sec{sec:coulomb}. 

To investigate the effect of rotation in a phenomenologically interesting case of charmonium one would have to consider rotation of the spherical symmetric potential, which has to be dealt with numerically. Nevertheless, we can make a ballpark estimate using \eq{g23}. The binding energy of $\jpsi$ is $M(\psi')-M(\jpsi)=\frac{16}{9}\alpha^2 M=0.59$~GeV. The reduced  mass is $M=m_c/2=0.63$~GeV which implies that $\alpha=0.72$.  Using the values $\ell_{1,0}$ and $\ell_{2,0}$ listed at the end of \sec{sec:coulomb} we derive that $\jpsi$ and $\psi'$ dissociate into the continuum when they get swirled by a vortex rotating with angular velocity $\Omega=3$~fm$^{-1}$ and $\Omega=0.6$~fm$^{-1}$ respectively. These estimates of $\Omega$ are somewhat larger than the vorticity achievable at a (relatively) low energy relativistic heavy ion collisions \cite{Deng:2016gyh}. We stress that at the above values of vorticity, the bound states completely dissolve, however the effect of rotation --- the decrease of the binding energy with $\Omega$ ---  is essential even at sub-critical angular velocities. Needless to say that these estimates should be taken only as a motivation for further investigation with more accurate quarkonium models.

We also computed the intensity of radiation emitted by a rotating bound state and found that when $R\Omega<1$ it increases by a factor $(R\Omega)^6$. This indicates an increase in transition probability from the excited to the ground state. 

Our results indicate that  plasma rotation has significant effect on the hadron spectra in relativistic heavy-ion which deserves a dedicated quantitative study.

%%%%%%%%%%%%%%%%%%%%%%%%%%%%%%%%
\acknowledgments
%I  am grateful to ... for many fruitful discussions of related problems. 
%We thank ... for helpful communications/correspondance.
This work  was supported in part by the U.S. Department of Energy under Grant No.\ DE-FG02-87ER40371.

%% appendix 
%\appendix
%\section{}\label{appA}

%%%%%%%%%%%%%%%%%%%%%%%%%%%%%%%%%%%%%

\end{document}